\documentstyle[prl,preprint,aps,epsfig]{revtex}
\begin{document}

\title{Effect of the 3K background radiation on ultrahigh energy cosmic rays}

\author{L. A. Anchordoqui, M. T. Dova, L. N. Epele}
\address{ Departamento de F\'\i sica, Universidad Nacional de La Plata\\
 C.C. 67, (1900) La Plata\\
\it Argentina\\ }
\author{and}   
\author{J. D. Swain}  
\address{ Department of Physics, Northeastern University\\
 Boston, Massachusetts, 02115\\ 
 USA}
\maketitle

\begin{abstract}

In this work we re-examine the opacity of the cosmic background
radiation to the propagation of extremely high energy cosmic rays. 
We use the continuous energy loss approximation to provide
spectral modification factors for several hypothesized cosmic ray
sources. Earlier problems with this approximation are
resolved including the effects of resonances other 
than the $\Delta$.

\end{abstract}
\hfill


\hfill

\begin{center}
{\em Accepted for publication in Physical Review D.}
\end{center}
\newpage

\section{Introduction}

With the discovery of the microwave background radiation (MBR)
\cite{Penzias}, it was realized that its effect on the 
propagation of ultrahigh energy cosmic rays (UHECR's) 
must be considered. The first treatments in 1966 by Greisen, Zatsepin 
and Kuz'min \cite{G,Z} indicated a sharp cutoff for cosmic rays
with energies above  \mbox{$ 5 \times 10^{19}$ eV,} due to 
the process $ \gamma + p  \rightarrow  \Delta  \rightarrow p/n  \pi$.
Physically, the thermal photons are seen highly blue-shifted by 
the cosmic rays in their rest frames so that this reaction
becomes possible.

Recently, several extensive air showers have been clearly observed which
imply the arrival of cosmic rays of energies above $10^{20}$ eV. 
In particular, the Akeno Giant Air Shower Array (AGASA) experiment recorded an event
with energy 1.7 - 2.6 $\times 10^{20}$ eV \cite{Yoshi,Hasha}, while 
the Fly's Eye experiment reported the highest energy cosmic ray (CR) event ever
detected on Earth, with an energy 2.3 - 4.1 $\times 10^{20}$ eV
\cite{Bird1,Bird2}. 
Since the GZK cutoff provides an important constraint on the  
proximity of UHECR sources, the origin of such energetic 
particles has became one of the most pressing
questions of cosmic ray astrophysics.

The study of the modification of 
the spectrum of cosmic rays due to their interactions 
with the MBR {\em en route} to us here on Earth, is essential if one
is to obtain a deeper understanding of the origin of the highest
energy cosmic rays.  
Various treatments of the propagation of nucleons
in the intergalactic medium exist, including purely 
analytical ones  \cite{BZ,S,BG,Stk,A,B},
as well as purely numerical ones \cite{Hill,Yoshida,Cronin,Lee}.
In this paper, we  re-analyze the effects of
meson photoproduction, the dominant energy loss mechanism,
using the continuous energy loss approximation which is expected to
be reasonable for the problem at hand. This allows us
to discuss, in a general way, the modification of the spectrum of
CR's from nearby sources due to interactions with the
cosmic background radiation.

\section{Propagation of Cosmic Rays in the Microwave Background Radiation}

The propagation of cosmic rays through the intergalactic medium is
described by a kinetic transport equation which takes into account
three mechanisms for energy losses: redshift, pair production, 
and pion photoproduction.
Nevertheless, for nearby sources corresponding
to propagation times $\tau \approx 3 \times 10^8$ years,
or distances of less than 100 Mpc, redshift
effects can be neglected. 

If one assumes that the highest energy
cosmic rays are indeed nucleons, the fractional  
energy loss due to interactions
with the cosmic background radiation at temperature $T$ and redshift
$z=0$, is determined by the integral of  the nucleon loss 
energy per collision times
the probability per unit time for a nucleon collision moving 
through an isotropic gas of photons \cite{Beri}. This integral can be
explicitly written as follows,
\begin{equation}
-\frac{1}{E} \frac{dE}{dt} =\frac{c}{2 \Gamma^2}\, \int_0^{w_m} dw_r \,\, 
K (w_r) \, \,  
\sigma (w_r)\, \, w_r \, \int_{w_r/2 \Gamma}^{w_m} dw \, 
\frac{n(w)}{w^2}   
\end{equation} 
where $w_r$ is the photon energy in the rest frame of the nucleon, and
$K(w_r)$ is the average fraction of the energy lost by
the photon to the nucleon in the laboratory frame (i.e. the frame
in which the microwave background radiation is at $\approx 3$ K).
n(w)dw stands for the number density
of photons with energy between w and dw, following a Planckian
distribution at temperature $T$ \cite{COBE}, $\sigma(w_r)$ is the total cross section of interaction, 
$\Gamma$ is the usual  
Lorentz factor of the nucleon,
and $w_m$ is the maximum energy of the photon
in the photon gas.

Thus, the fractional energy loss is given by
\begin{equation}
-\frac{1}{E} \frac{dE}{dt} = - \frac{ckT}{2 \pi^2 \Gamma^2 (c \hbar)^3}
\int_{w_0}^{\infty}  dw_r \,
\sigma (w_r) \,K (w_r)\, w_r \, \ln ( 1 -  e^{-w_r / 2 \Gamma kT})
\label{!}
\end{equation}
where $k$ and $\hbar$ are Boltzmann's and Planck's constants
respectively, and $w_0$ is the threshold energy for the reaction 
in the rest frame of the nucleon.

The characteristic time for the energy loss due to pair production at 
$E > 10^{19}$ eV is $t \approx 5\times 10^9$ yr \cite{BLUE}
and therefore it does not affect the spectrum of nucleons arriving from
nearby sources. Consequently, for nucleons with 
$E > 3 \times 10^{19}$eV (taking into account the interaction
with the tail of the Planck distribution), meson photoproduction is the 
dominant mechanism for energy loss. Notice that we do not 
distinguish between neutrons and protons; in addition, the inelasticity due 
to the neutron $\beta$ decay is negligible. 

In order to determine the effect of meson photoproduction on the
spectrum of cosmic rays, we first examine the kinematics of
photon-nucleon interactions. Assuming that reactions mediated 
by baryon resonances have spherically symmetric decay angular distributions 
\cite{B,Cronin}, the average energy loss of the nucleon after n resonant 
collisions is given by
\begin{equation}
K(m_{R_0}) = 1 - \frac{1}{2^n} \prod_{i=1}^{n} \left( 1 + \frac{m_{R_{_i}}^2 -
m_M^2}{m_{R_{_{i-1}}}^2} \right)
\end{equation}
where $m_{R_{_i}}$ denotes the mass of the
resonant system of the chain,  $m_M$ the mass of the associated
meson, $m_{R_{_0}} = \sqrt{s}$ is the total energy of the reaction 
in the centre
of mass, and $m_{R_{_n}}$ the mass of the nucleon.
It is well established from experiments \cite{inel} that, at very 
high energies 
($\sqrt{s}$ above $\sim 3$ GeV), the incident nucleons
lose one-half their energy via pion photoproduction independent of the
number of pions produced. 

In the region dominated by baryon resonances, the cross section
is described by a sum of Breit-Wigner distributions constructed  
from the experimental data in the Table of Particle Properties \cite{PDG}
considering the main resonances produced in
$N \gamma$ collisions with $\pi N, \pi \pi N,\, {\mathrm{and}} \, \pi K N$ 
final states. For the cross section at high energies we used 
the fits to the 
high-energy cross section $\sigma_{{\mathrm{total}}} (p\gamma)$ 
made by the CERN, DESY HERA, and COMPAS Groups \cite{PDG}.
In this energy range, the $\sigma_{{\mathrm{total}}} (n\gamma)$ is 
to a good approximation identical to $\sigma_{{\mathrm{total}}} (p\gamma)$. 

The numerical integration of Eq. (\ref{!}) is performed taking 
into account the
aforementioned resonance decays and the production of multipion final
states at higher centre of mass energies ($\sqrt{s} \sim 3$ GeV). 
A $\chi^2$ fit of the numerical results of equation (\ref{!}) 
with an exponential behaviour, $ A \, {\mathrm{exp}} \{ - B / E \}$, 
proposed 
by Berezinsky and Grigor'eva \cite{BG} for
the region of resonances, gives: $ A = ( 3.66  \pm 0.08 )\times 10^{-8} \, {\mathrm{yr}}^{-1} $, $B =
(2.87 \pm 0.03 )\times
10^{20}$ eV with a  $\chi^2/dof = 3.9/10$. 
At high energies the 
fractional energy loss was fitted with a constant, $ C = ( 2.42 \pm 0.03 ) \times 10^{-8}  \,\, 
{\mathrm{yr}}^{-1} $.
These results differ from those obtained in \cite{BG}, due to a 
refined 
expression for the total cross section. From the values determined 
for the fractional energy loss it is straightforward to
compute the energy 
degradation of UHECR's in terms of their flight distance. 
The results shown in Fig. 1, are consistent with those previously obtained by 
Cronin \cite{JCronin}, in which 
just single pion production is considered. Our results therefore support
the assumption that the energy loss is dominated
by single pion photoproduction.

\section{Modification of the Cosmic Ray Spectrum}

Let us now turn to the modification that the MBR produces in the 
UHECR spectrum. The evolution of
the spectrum is governed by the balance equation 
\begin{equation}
\frac{\partial N}{\partial t} = \frac{\partial (b(E) N)}
{\partial E} + D \,\, \nabla^2 N + Q
\label{@}
\end{equation}
that takes into account the conservation of the
total number of nucleons in the spectrum.
In the first term  on the right, b(E) is the mean rate at which particles 
lose
energy. 
The second term, the diffusion in the MBR,
is found to be extremely small due to the low density of relic
photons and the
fact that the average cosmic magnetic field is less than  $10^{-9}$ G 
\cite{mgfield} 
and is neglected in the following.
The third term corresponds to the particle injection rate into
the intergalactic medium from some hypothetical source.
Since the origin of cosmic rays is still unknown, we consider
three possible models:
1) the universal hypothesis, which assumes that cosmic rays come from
no well-defined source, but rather
are produced uniformly throughout space, 2) single point sources
of cosmic rays, and 3) sources of finite size
approximating clusters of galaxies.
In all the cases it has been assumed that nucleons propagate
in a straight line through the intergalactic medium due to the reasons 
mentioned above that allow us to neglect the diffusion term.

There exists evidence that
the source spectrum of cosmic rays has a power-law dependence
$ Q (E) = \kappa \, E^{- \gamma}$ (see for instance \cite{Hill}). 
With the hypothesis that cosmic rays are produced from sources located
uniformly in space, with this power law energy dependence (which
implies a steady state process), the solution of Eq. (\ref{@}) is
found to be 
\begin{equation} 
N(E) = \frac{Q(E) \, E }{ b(E)\, (\gamma - 1)}
\end{equation} 

For the case of a single point source,
the solution 
of equation (\ref{@}) reads \cite{Stk},
\begin{equation}
N(E, t) = \frac{1}{b(E)} \, \int_E^{\infty} \, Q(E_g, t') \,dE_g 
\label{spectrum}
\end{equation}
with
\begin{equation}
t'=  t - \int_E^{E_g} \frac{d\tilde{E}}{b(\tilde{E})}
\label{espectro}
\end{equation}
and $E_g$ the energy of the nucleon when emitted by the source.
The injection spectrum of a single  source located at $t_0$ from the observer
can be written as $Q(E, t) \,= \,\kappa \, E^{- \gamma} \,
\delta(t - t_0)$ and for simplicity, we consider the distance as measured
from the source, that means $t_0 = 0 $. 
At very
high energies, i.e. where $b(E) = C \, E$ (where $C$ is the constant
defined above in the discussion of fractional energy loss) 
the total number of particles at a given distance from the source is 
given by
\begin{equation}
N(E, t) \approx \frac{\kappa}{b(E)} \int_E^{\infty}\, E_g^{-\gamma} \,
\, \, \delta\left( t - \frac{1}{C} \ln 
\frac{E_g}{E} \right) dE_g
\end{equation}
or equivalently
\begin{equation}
N(E, t) \approx \kappa \, E^{-\gamma} e^{- \, (\gamma -1) \, C \, t }
\end{equation}
This means that the spectrum is uniformly damped by a factor depending
on the proximity of the source.

At low energies, in the region dominated by baryon resonances, the
parametrization of $b(E)$ does not allow a complete analytical solution.
However using the change of variables 
\begin{equation}
{\tilde t} =
\int_E^{E_g} \frac{d\tilde{E}}{ b(\tilde{E})}
\end{equation}
with $E_g = \xi (E, \tilde{t})$ and $d{\tilde t} =
dE_g/b(E_g)$, we easily obtain,
\begin{equation}
N(E, t) =  \frac{\kappa}{b(E)} \int_0^{\infty}\, \xi(E, \tilde{t})^
{-\gamma}\,
\,\delta({\tilde t} - t) \,\, b [ \xi(E, {\tilde t}) ] \,\,d{\tilde t}
\end{equation}
and then, the compact form,
\begin{equation}
N(E, t) = \frac{\kappa}{b(E)} E_g^{-\gamma} b(E_g)
\end{equation}
In our case, where we have assumed an exponential behaviour of 
the fractional loss energy, $E_g$ is fixed by the constraint:  
$\,
A \, t \, - \,  {\mathrm{Ei}}\,(B/E) 
+ \, {\mathrm{Ei}}\, (B/E_g) 
= 0
$,  Ei being the exponential integral \cite{Abra},
and $B$ the constant defined above in the parametrization of
Berezinsky and Grigor'eva.

Studies are underway of the case of a nearby source 
(i.e. within about 100 Mpc)
for which 
the fractional energy loss $\Delta E/E$ is small. Writing $E_g =
E + \Delta E$, and neglecting higher orders 
in $\Delta E$,
it is possible to obtain an analytical solution
and the most relevant characteristics of the modified spectrum.
Therefore, an expansion of
$b(E + \Delta E)$ in powers of $\Delta E$ in Eq. (\ref{espectro}) 
allows one to obtain,
an expression for
$\Delta E$,
\begin{equation}
\Delta E \approx \left\{ {\mathrm{exp}}\left[ \frac{t \, b(E)\,
(E + B)}{E^2} \right] - 1 \right\}
\frac{E^2}{(E + B)} 
\label{l1}
\end{equation}

To describe the modification of the spectrum, 
it is convenient to 
introduce the factor $\eta$, the ratio between the
modified spectrum and the unmodified one, that results in 
\begin{equation}
\eta = \left( \frac{E + \Delta E}{E}\right)^{-\gamma} 
\frac{b(E+\Delta E)}{b(E)}
\label{l2}
\end{equation}  
Equations (\ref{l1}) and (\ref{l2}) describe the spectral modification
factor up to energies of $\approx 95$ EeV ($\approx
85$ EeV) with a precision of
4\% (9\%) for a source situated at 50 Mpc (100 Mpc).

In Fig. 2 we plot the modification factors for the case of 
nearby sources with power law injection ($\gamma = 2.5$) to 
compare with the corresponding results of ref. \cite{BG}. The most 
significant features of $\eta$ are the 
bump and the cutoff. It has been
noted in the literature that the continuous energy loss approximation
tends to overestimate the size of the bump \cite{A,Lee}. Figure 2 shows 
how the bump is less pronounced in our treatment. 
As the bump is a consequence of the
sharp (exponential) dependence of the free path of the nucleon on energy, 
we attribute these differences to the
replacement of a cross section 
approximated by
the values at threshold energy \cite{BG}, by a more detailed expression
taking into account the main baryon resonances that turn out to be important.

Another alternative for the particle injection
is clusters of galaxies. These are usually modelled by a set of point sources with 
spatially uniform distribution, although in reality the 
distribution of galaxies inside the clusters is somewhat non-uniform. 
In our treatment we shall assume that the concentration of potential
sources at the center of
the cluster is higher than that 
near the periphery, and we adopt a spatial 
gaussian distribution.
With this hypothesis, the particle injection rate into the
intergalactic medium is given by  
\begin{equation}
Q(E,t) = \kappa \int_{-\infty}^{\infty} \frac{E^{-\gamma}}{\sqrt{2 \, \pi}\, 
\sigma} \, \delta(t-T) \; 
{\mathrm{exp}}\left\{ \frac{-(T-t_0)^2}{2\,\sigma^2} \right\} dT
\end{equation}
A delta function expansion around  $t_0$, with derivatives denoted
by lower case Roman superscripts,
\begin{equation}
\delta(t-T) = \delta(t-t_0) + \delta^{(i)}(t-t_0) (T-t_0)+\frac{1}{2!}
\delta^{(ii)}(t-t_0) (T-t_0)^2 + \dots
\end{equation}
leads to a convenient form for the injection spectrum, which is given
by,
\begin{equation}
Q(E,t) =  \kappa E^{-\gamma} \,[\,\delta(t^{\prime}-t_0) + 
\frac{\sigma^2}{2!} \, 
\delta^{(ii)}(t^{\prime}-t_0) + \frac{\sigma^4}{4!}\, \delta^{(iv)}(t^{\prime}-t_0)
+\dots]
\end{equation}
From Eqs. (\ref{spectrum}) and (\ref{espectro}), it is
straightforward to compute an expression for the modification factor,
\begin{equation}
\eta = \frac{E_g^{-\gamma}\, b(E_g)}{E^{-\gamma} \, b(E)} \, 
 \left\{ 1+\frac{ \sigma^2 A^2 e^{-2B/E_g}}{2!} F_1(E_g) + \frac{\sigma^4 A^4 e^{-4B/E_g}}
{4!} F_2(E_g)  + {\cal O}(6) \right\}
\end{equation}
where
\begin{displaymath}
F_1(E_g)= 2 B^2 E_g^{-2} + 
(2-3\gamma) B E_g^{-1} +(1-\gamma)^2
\end{displaymath}
and
\begin{eqnarray}
F_2(E_g) & = & 24 B^4 E_g^{-4} + (4 - 50
\gamma) B^3 E_g^{-3} + (35 \gamma^2 -25 \gamma +8) B^2 E_g^{-2}
\nonumber \\
& + & (-10
\gamma^3 +20 \gamma^2 - 15 \gamma +4) B E_g^{-1} + (1- \gamma)^4 \nonumber
\end{eqnarray} 

The modification factor for the case of extended sources
described by Gaussian distributions of widths
2 and 6 Mpc
at a distance of 18.3 Mpc is shown in Fig. 3. 
These may be taken as very crude models of the Virgo cluster \cite{Virgo},
assuming that there is no other significant energy loss mechanism
for cosmic rays traversing parts of the cluster beyond those due
to interactions with the cosmic background radiation.

A power law injection ($\gamma = 2.5$) was used again, as in the pointlike
case.
The curves can be understood qualitatively. Both peaks are in about
the same place, occurring around the onset of pion photoproduction.
The narrower curve, corresponding to the broader source distribution,
reflects the losses suffered by the more remote part of the distribution
in traversing a greater distance to us. We note that the modification factors
for nearby extended sources can appear similar to those for more distant
narrower sources.

\section{Conclusions}

We have presented a recalculation of the effects of the cosmic
background radiation on the propagation of cosmic rays using
the continuous energy loss approximation first used by
Berezinsky \cite{Beri} and our best knowledge
of $\gamma$-nucleon cross sections. This approximation is
important as it is the only way to allow semi-analytical 
treatments in a wide range of situations. 

While earlier calculations within this approximation
suffered from problems described in the main text, these
do not appear with our improved treatment of 
cross sections, and the results are in 
reasonable agreement with previous calculations using
Monte Carlo methods without the continuous energy loss
approximation.

The eventual observation of the GZK cutoff is of fundamental
interest in cosmic ray physics, providing constraints
on the distance to sources of ultrahigh enery cosmic rays.
The future Pierre Auger Project \cite{Desrep} should provide
a decisive test of the ideas discussed in this paper.

\section*{Acknowledgments}

\noindent We are grateful to Huner Fanchiotti and Carlos Garc\'{\i}a Canal 
for illuminating discussions, and to James Cronin for stimulating our
initial interest in this subject. We would also like to thank
Carlos Feinstein for helpful information on the morphology of the
Virgo cluster. This work was partially supported by CONICET, 
Argentina. L.A.A. thanks FOMEC for financial support.

\newpage

\centerline{Figure Captions}

\vspace{2.0cm}

Figure 1. Energy attenuation length of nucleons in the intergalactic medium. 
Notice that, independently of the initial energy of the nucleon, 
the mean energy values 
approach to 100 EeV after a distance of $\approx 100$ Mpc.

Figure 2. Modification factor of single-source energy 
espectra for different values of propagation distance and 
power law index $\gamma = 2.5$.

Figure 3. Modification factor of extended-source energy 
spectrum for a propagation distance $\approx$ 18.3 Mpc and 
power law index $\gamma = 2.5$. Solid line stands for a 
gaussian distribution of width $\sigma = 2$ Mpc, 
while dashed line a width of $\sigma = 6$ Mpc.

\clearpage


\begin{figure}[htbp]
\begin{center}
\epsfig{file=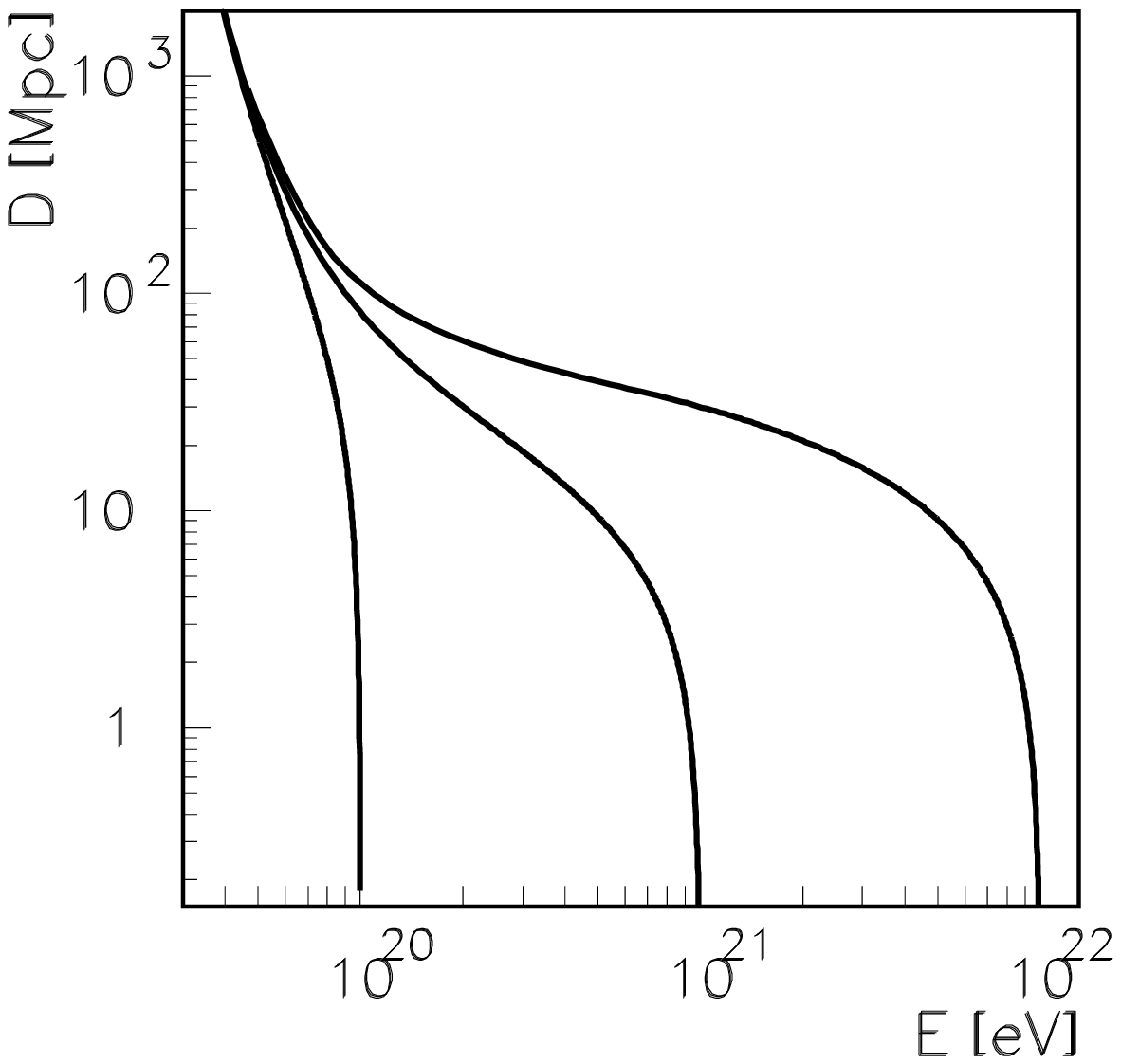,width=20cm,clip=}
\end{center}
\vspace{0.1cm}
\caption{\ }
\end{figure}

\clearpage

\begin{figure}[htbp]
\begin{center}
\epsfig{file=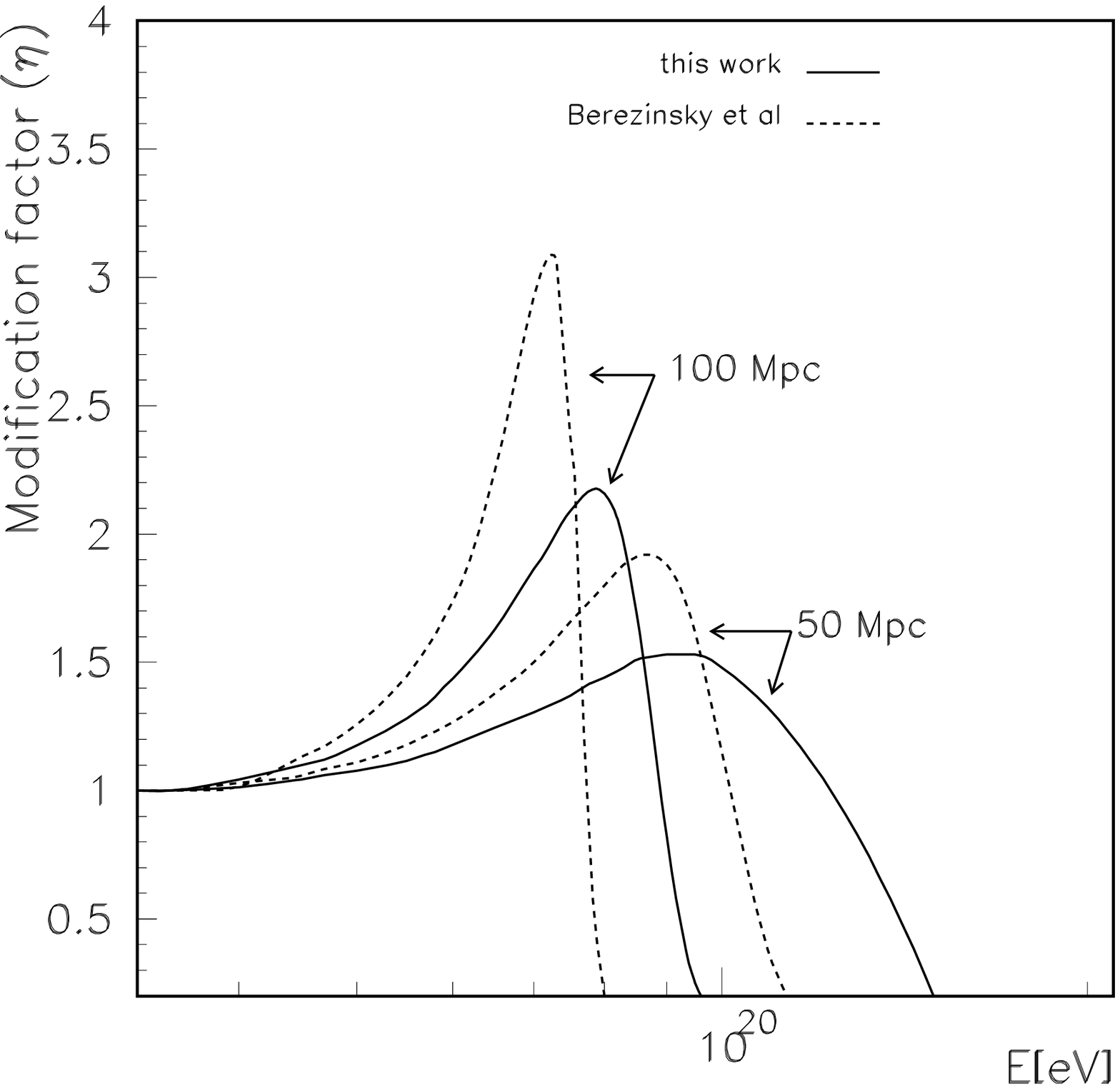,width=20cm}
\end{center}
\vspace{1cm}
\caption{\ }
\end{figure}

\clearpage

\begin{figure}[htbp]
\begin{center}
\epsfig{file=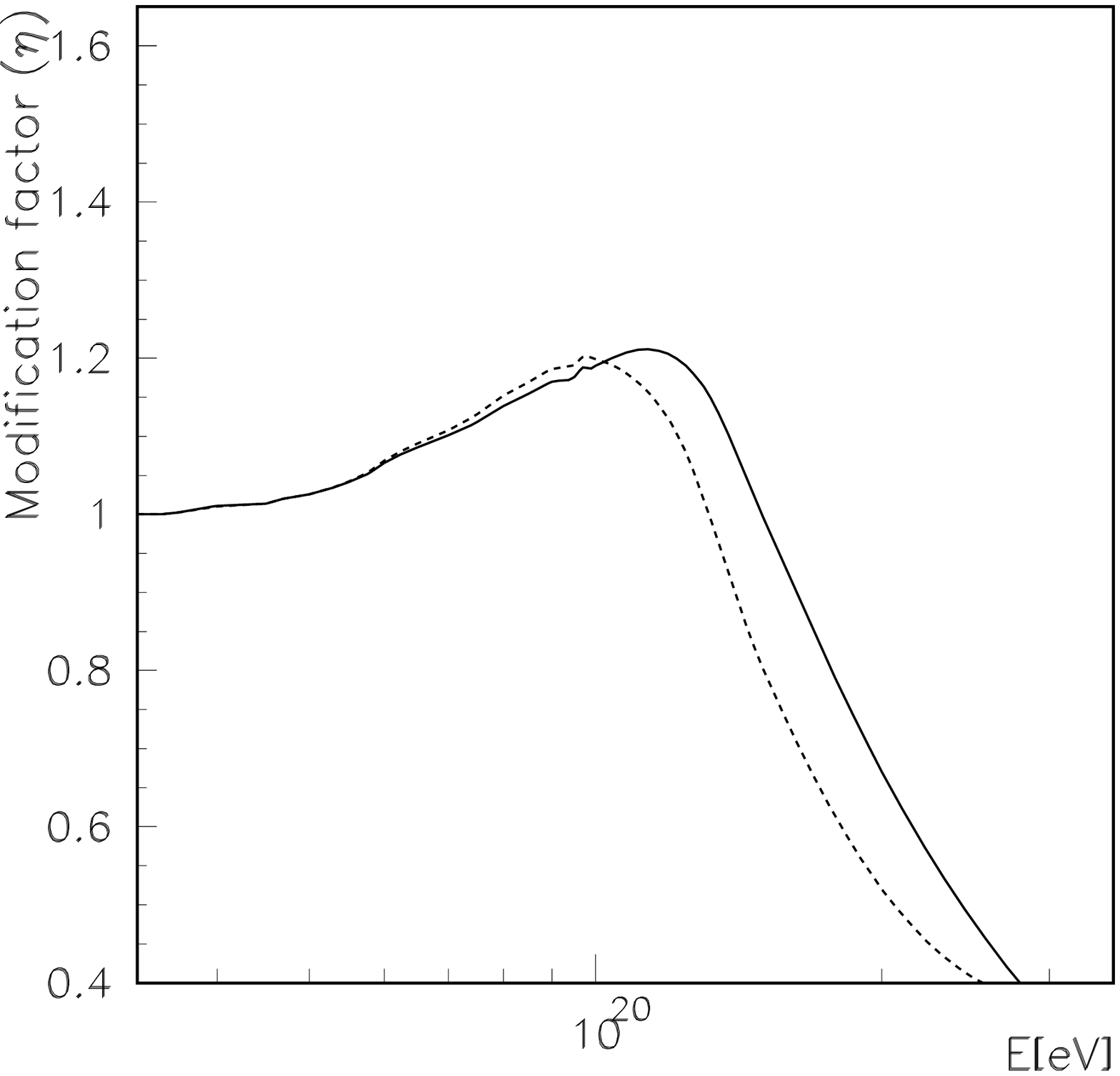,width=20cm}
\end{center}
\vspace{1cm}
\caption{\ }
\end{figure}


\newpage

\begin{thebibliography}{99}
\bibitem{Penzias} A. A. Penzias and R. W. Wilson, {\em Ap. J} {\bf 142},
419 (1965).
\bibitem{G} K. Greisen, {\em Phys. Rev. Lett.} {\bf 16}, 748 (1966).
\bibitem{Z} G. T. Zatsepin and V. A. Kuz'min, {\em Pis'ma Zh. Eksp.
Teor. Fiz.} {\bf 4}, 114 (1966).
\bibitem{Yoshi}S. Yoshida {\em et al}., {\em Astropart. Phys.} {\bf3}, 105
(1995).
\bibitem{Hasha} N. Hashashida {\em et al}., {\em Phys. Rev. Lett.} {\bf
73}, 3491 (1994).
\bibitem{Bird1} D. J. Bird {\em et al}., {\em Phys. Rev. Lett.} {\bf 71},
3401 (1993). 
\bibitem{Bird2} D. J. Bird {\em et al}., {\em Ap. J.} {\bf 441}, 144 (1995).
\bibitem{BZ} V. S. Berenzinsky and G. T. Zatsepin, {\em Yad. Fiz.} {\bf 13}, 
797 (1971).
\bibitem{S} F. W. Stecker, {\em Phys. Rev. Lett.} {\bf 21}, 1016 (1968).
\bibitem{BG} V. S. Berezinsky and S. I. Grigor'eva, 
{\em Astron. Astrophys.} {\bf 199}, 1 (1988).
\bibitem{Stk} F. W. Stecker, {\em Nature} {\bf 342}, 401 (1989).
\bibitem{A} F. A. Aharonian, B. L. Kanevsky and V. V. Vardanian, {\em Astrophys. 
Space Sci.} {\bf 167}, 93 (1990).
\bibitem{B} J. P. Rachen and P. L. Biermann, {\em Astron. Astrophys.} {\bf 272},
161 (1993).
\bibitem {Hill} C. T. Hill and D. N. Schramm, {\em Phys. Rev.} {\bf D31}, 
564 (1985).
\bibitem {Yoshida} S. Yoshida and M. Teshima, {\em Prog. Theor. Phys.} {\bf 89},
833 (1993).
\bibitem {Cronin} F. A. Aharonian and J. W. Cronin, {\em Phys. Rev.} {\bf D50}, 
1892 (1994).
\bibitem{Lee} S. Lee, Report No. FERMILAB-Pub-96/066-A (unpublished).
\bibitem{Beri} V. S. Berenzinsky,  {\em Yad. Fiz.} {\bf 11}, 399 (1970).
\bibitem{COBE} Concerning the density of photons, data from the Far-InfraRed Absolute 
Spectrometer (FIRAS) 
of the Cosmic Background Explorer (COBE) shows that
the spectrum of MBR is that of a blackbody of temperature $T = 2.73
\pm 0.06$ K, with no deviation from a Planckian spectrum greater than
0.25 \% of the peak brightness and  a limit on the Compton $y$-parameter 
of $y < 0.0004$ (95 \% CL). See, R. A. Shafer {\em et al}., 
{\em Bull. Am. Phys. Soc} {\bf 36}, 1398 (1991); H. P. Gush, 
M. Halpern and E. H. Wishnow, {\em Phys. Rev. Lett.} {\bf 65}, 537 (1990).  
\bibitem{BLUE} G. Blumenthal, {\em Phys. Rev.} D {\bf 1}, 1596 (1970).
\bibitem{inel} I. Golyak, {\em Mod. Phys. Lett.}  {\bf A7}, 2401 (1992), 
and references therein.
\bibitem{PDG} Particle Data Group, L. Montanet {\em et al}., {\em Phys. Rev.} 
{\bf D50}, 1173, 1335 (1994).
\bibitem{JCronin} J. W. Cronin, {\em in Cosmic Rays Above 10$^{19}$
eV}, Proceedings of the International Workshop, Paris, France, 1992,
edited by J. Cronin, A. A. Watson, and M. Boratav [Nucl. Phys. B 
(Proc. Suppl.) {\bf 28B}, 213 (1993)]. See also reference \cite{Cronin}.
\bibitem{mgfield} P. P. Kronberg, {\em Rep. Prog. Phys.} {\bf 57},
325 (1994)
\bibitem{Abra} M. Abramowitz and I. A. Stegun, ``Handbook of
Mathematical Functions'' (Dover, New York, 1970).
\bibitem{Virgo} B. Binggeli, G. A. Tammann and A. Sandage, {\em Astron. J.} 
{\bf 94}, 251 (1987).
\bibitem{Desrep} The Auger Collaboration, Pierre Auger Project Design Report, 
1995, Fermi National Accelerator Laboratory (unpublished).
\end{thebibliography}
\end{document}